\documentclass[twocolumn,aps,prl,10pt,amsmath,amssymb,nofootinbib,showpacs,superscriptaddress,floatfix]{revtex4-1}

\DeclareFontFamily{U}{rcjhbltx}{}
\DeclareFontShape{U}{rcjhbltx}{m}{n}{<->rcjhbltx}{}
\DeclareSymbolFont{hebrewletters}{U}{rcjhbltx}{m}{n}

\usepackage{graphicx}
\usepackage{color}
\usepackage[usenames,dvipsnames]{xcolor}
\usepackage[colorlinks=true,linkcolor=Red,citecolor=Green,linktoc=page]{hyperref}
\usepackage{multirow}
\usepackage{float}
\usepackage{flushend}
\usepackage{balance}
\usepackage[varg]{txfonts}
\usepackage{ulem}
\usepackage{fancyhdr}

\DeclareMathSymbol{\lamed}{\mathord}{hebrewletters}{108}

\begin{document}
\title{Bulk superinsulation and polar nematic order in nanopatterned NbTiN}

\author{A.\,Yu.\,Mironov}

\affiliation{A. V. Rzhanov Institute of Semiconductor Physics, 630090 Novosibirsk, Russia}

\author{C.\,A.\,Trugenberger}

\affiliation{SwissScientific Technologies SA, rue du Rhone 59, CH-1204 Geneva, Switzerland}
\affiliation{Division of Science, New York University Abu Dhabi, Abu Dhabi, United Arab Emirates}

\author{M.\,C.\,Diamantini}

\affiliation{NiPS Laboratory, INFN and Dipartimento di Fisica e Geologia, University of Perugia, via A. Pascoli, I-06100 Perugia, Italy}

\author{D.\,A.\,Nasimov}

\affiliation{A. V. Rzhanov Institute of Semiconductor Physics, 630090 Novosibirsk, Russia}

\author{V. M. Vinokur}
\affiliation{Terra Quantum AG, Kornhausstrasse 25, CH-9000 St. Gallen, Switzerland}

\
%\date{\today}
\begin{abstract}
We present an experimental evidence of 3D superinsulation in a nanopatterned slab of NbTiN, given by the Vogel-Fulcher-Tamman (VFT) scaling of the conductance when approaching the critical temperature from above and by the vanishing of the conductance below the transition. In the electric Meissner state, we find polar nematic order arising from ferroelectric alignement of short electric strings excited by external electromagnetic fields. Our results prove that superinsulation appears also in ordered structures provided that these are large enough, thereby confirming the origin of  superinsulation as electric confinement, independent of disorder.
\end{abstract}
\maketitle

\section*{Introduction}~~

Superinsulators\,\cite{dst, vinokurnat, confinement}, see\,\cite{book} for a review, are materials hosting a macroscopic quantum state which is a mirror image of superconductivity.
In superinsulators, the effective electromagnetic action is compact quantum electrodynamics (QED) at strong coupling\,\cite{polyakov}. It is the compactness of the gauge group that entails the presence of magnetic monopoles, see\,\cite{goddardolive} for a review, which are instantons\,\cite{coleman} in two spatial dimensions (2D) and particles in three spatial dimensions (3D). Correspondingly, the superinsulating state is a plasma of (non-relativistic) magnetic monopole instantons in 2D\,\cite{confinement} and a magnetic monopole condensate in 3D\,\cite{moncond}. In these states, electric fields are constrained into flux tubes between Cooper pairs and Cooper holes. These flux tubes are dual to Abrikosov vortices in superconductors, with the difference that ``electric vortices" can end on internal sources. An elementary excitation made of an electric flux tube connecting a Cooper pair and a Cooper hole is the purely electric equivalent of a strong interaction pion, with Cooper pairs playing the role of quarks. The flux tubes are confining strings preventing the separation of charges and leading to an infinite resistance, even at finite temperatures\,\cite{confinement, moncond}, for the same reason that quarks can never be extracted from pions and other hadrons. Hence the name ``superinsulators".  Magnetic monopoles and confining strings, which were just abstract objects allowed in compact versions of quantum electrodynamics (QED)\,\cite{polyakov} become experimentally observable physical objects in superinsulators\,\,\cite{dst, vinokurnat, confinement, book}.

Superinsulation is destroyed when the string tension vanishes at either sufficiently high temperature or at large applied voltage. The corresponding disappearance of superinsulation is a phase transition characterized by the temperature scaling of the resistance. These temperaure scalings are different in two-dimensional (2D) and 3D (bulk) superinsulators. In a 2D superinsulator, the temperature scaling assumes the classical Berezinskii-Kosterlitz-Thouless (BKT) form\,\cite{ber, kt1, kt2}, whereas in 3D it changes to a Vogel-Fulcher-Tamman (VFT) scaling\,\cite{vft}, exactly as for the dual type-III superconductors\,\cite{typeIII}.

Superinsulators have been detected in  superconducting films of various materials\,\cite{shahar, vinokurnat, mironov1, pion, mironov2}, one of which was thick enough to show hints of a 3D behaviour\,\cite{shahar}. All these films showing superinsulation appear to have an emergent granular structure\,\cite{granular1, granular2}: they are arrays of superconducting granules connected by Josephson tunnelling junctions. This implies that these materials are effective Josephson junction arrays (JJA), see\,\cite{jja} for a review, which are the metamaterial for which superinsulation has been first predicted\,\cite{dst}. In all these films the superinsulating state was obtained by driving them across the superconductor-to-insulator transition (SIT)\,\cite{vinokurnat}, see\,\cite{book} for a review, by decreasing the films thickness, or by applying a magnetic field or a gate voltage.

Here we report the observations obtained for a material which is by itself a bulk (3D) superinsulator at low temperatures, so that there is no need to drive it across a transition from a superconductor in order to observe superinsulation. This superinsulating metamaterial is obtained by nanopatterning a slab of NbTiN by punching a regular array of holes into it. As we now show, this is the first bulk superinsulator at temperatures below a critical temperature of $0.31 ^\circ {\rm K}$, which is more than five times higher than the typical critical temperature in superinsulating films.

Superinsulation has not yet been observed in JJA. As was pointed out in\,\cite{poccia}, this is happening since, in presently available arrays, having typically about 500 to 1000 islands in their very largest realizations, it is difficult to simultaneously obtain large enough electric screening lengths due to the islands' self-capacitances and accomodate an entire electric string. The structure presented in this paper is essentially a regular JJA with 10 million islands and a very large electric screening length due to the large dielectric constant of NbTiN. As originally predicted\,\cite{dst}, superinsulation immediately appears in such a large regular structure. If disorder would be the root cause of the diverging resistance of superinsulators, this phenomenon should not be observed in the present regular and ordered nanopatterned materials. The fact that, instead, it is clearly observed in these regular and ordered systems is the final proof that superinsulation has nothing to do with disorder.

Electric fields that arise from voltages below a critical voltage value do not penetrate superinsulators. This is a phenomenon dual to the usual Meissner effect for magnetic fields in superconductors, see\,\cite{pion, book} for a review. However, when sufficient energy is provided by external electric and magnetic fields, electric pions, which are thin strings of electric flux with $\pm$ charges at their ends, can be excited in the interior of the bulk sample. These electric strings move under the influence of the external magnetic field in the plane perpendicular to the magnetic field and arrange themselves under the influence of their interactions and the applied voltage in such a way that local electric polarization configurations arise in this plane. We show how, in  absence of external electric fields, polar nematic orders, see \,\cite{polar} for a recent review, arise as a consequence of the competition between interactions of electric strings and the effect of a magnetic field.  When an in-plane electric field is added, the induced polar orders become more complex. We observe the resulting effects by measuring configurations of the local, magnetic- and electric-field-induced voltages arising at various locations in the sample and in different directions. These configurations form only for ranges of applied magnetic field and voltage where the superinsulating state exists and electric pions can be excited, thus confirming their origin.

\section*{The nanopatterned materials}~~

\begin{figure}
	\begin{center}
		\includegraphics[width=\linewidth]{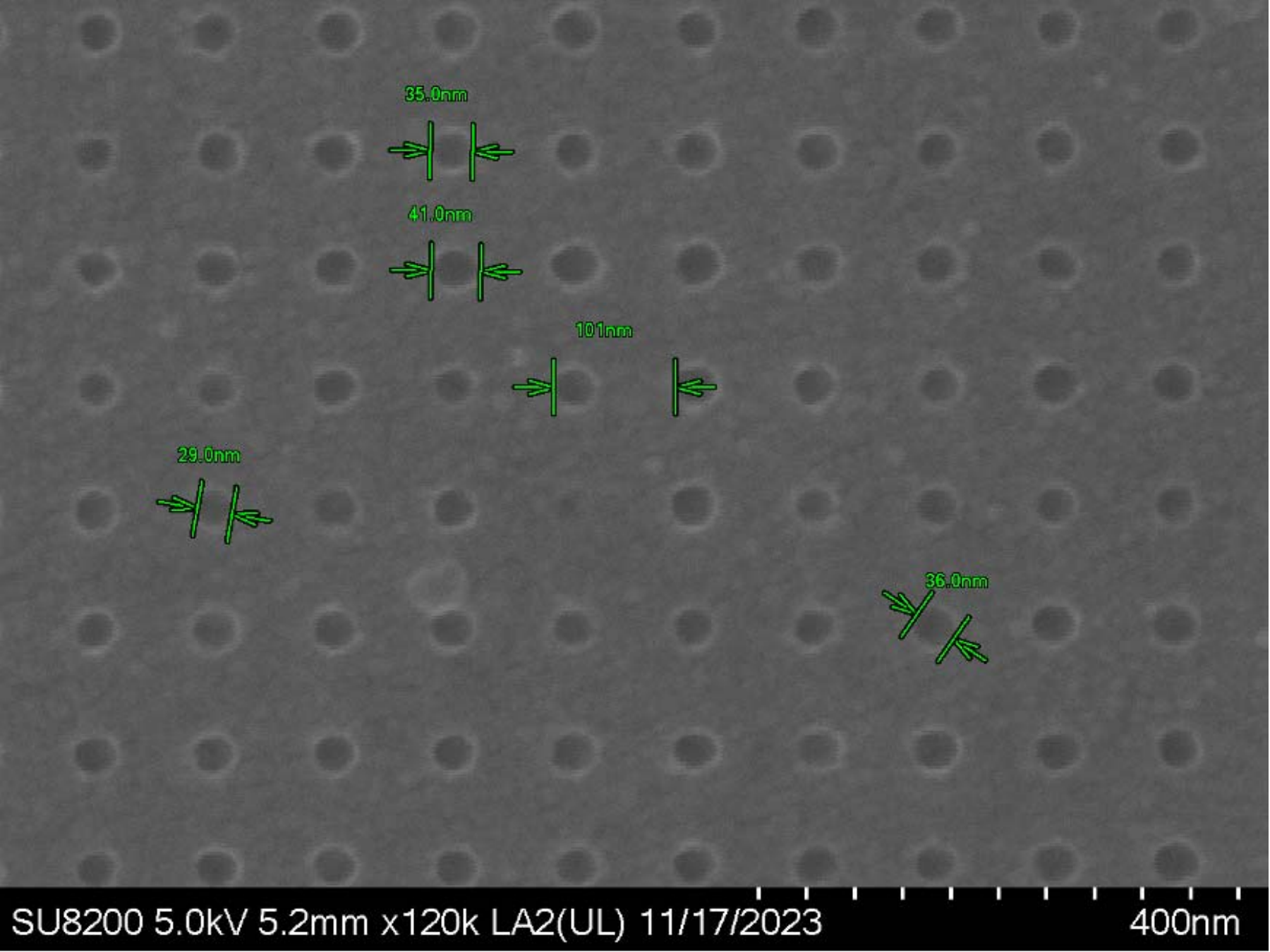}	
		\caption{\textbf{Perforated film.} Image of a section of one of the perforated NbTiN films, collected by scanning electron microscope.} \label{fig_perf}
	\end{center}
\end{figure}

The NbTiN films, having the thickness of 10\,nm, are grown using atomic layer deposition (ALD) on a SiO$_2$/Si substrate at a temperature $350^\circ$.  The structure of the films is of a polycrystalline nature with the grain size being about 5\,nm, as determined with high resolution electron microscopy.  To carry out the transport measurements, the films are first patterned by means of a conventional UV lithography and plasma etching into bridges of 100\,$\mu$m wide and with 100\,$\mu$m and 250\,$\mu$m distances between the voltage probes. Then, making use of electron lithography and subsequent plasma etching, a square lattice of holes with the diameter $\approx40$\,nm and the period $a=100$\,nm is created, covering an area $100\times 2000\,\mu$m$^2$ for the first film, see Fig.\,\ref{fig_perf}, and an area $100\times 1000\,\mu$m$^2$ for the second film. This corresponds to 10, or 5 millions lattice units for the whole surface of the sample, respectively. The experiment  is performed in the dilution refrigerators. %at the Institute of Semiconductor Physics in Novosibirsk.
The measurements of the temperature $T$ and the magnetic field $B$ dependencies of the resistance and differential resistance are carried out using a standard two-probe low-frequency ac/dc technique at the frequency 3\,Hz with an ac voltage 100\,nV. For all measurements we use lock-in amplifiers SR865a and sub-femtoamp\`eremeter Keythley 6430 and nanovoltmeter Agilent 34420a. The magnetic field is applied perpendicular to the film surface.

\section*{Discussion of experimental results}~~

\begin{figure}
	\begin{center}
		\includegraphics[width=\linewidth]{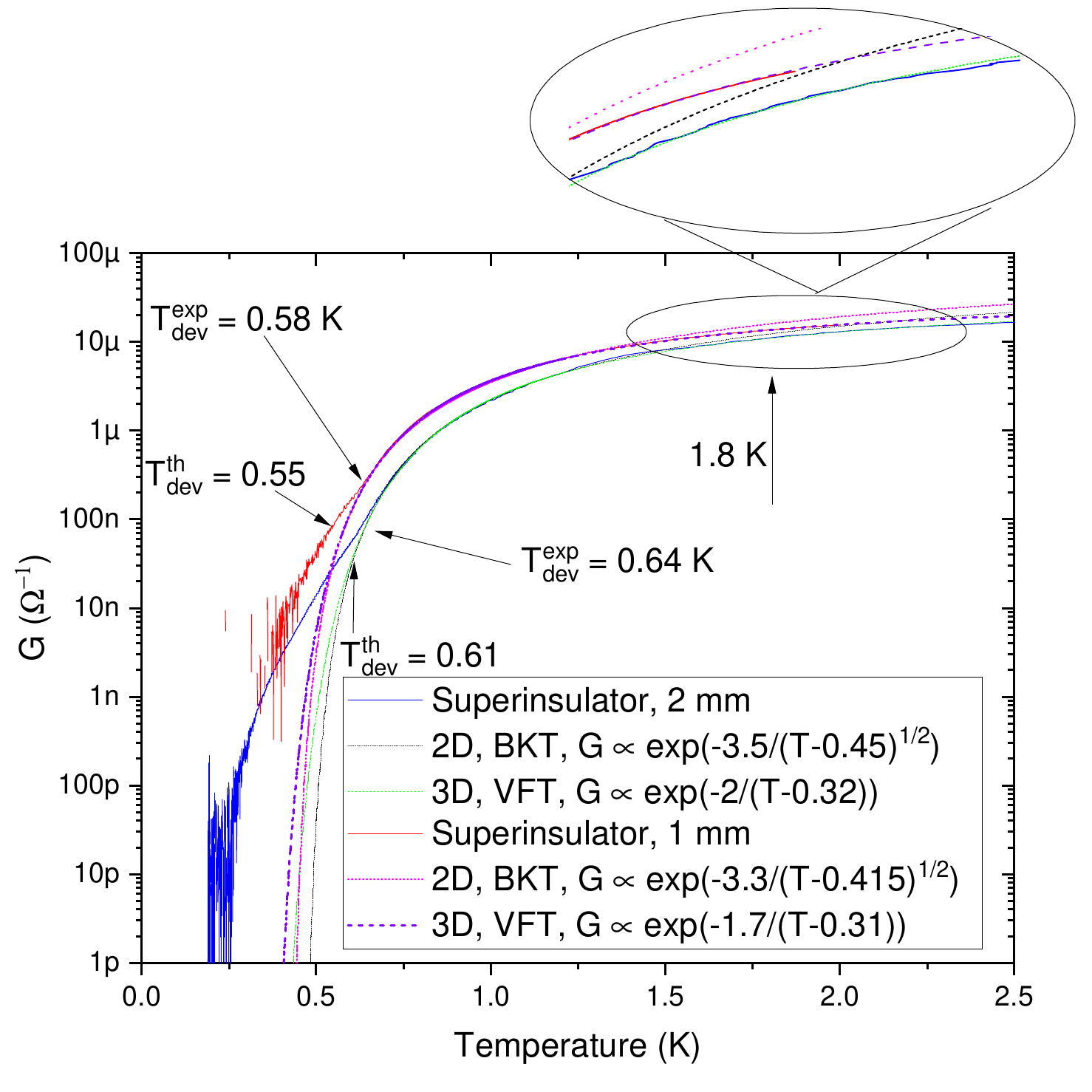}	
		\caption{\textbf{Temperature dependence of the conductance.} Temperature dependence of the conductances of the two nanoperforated films and fits by BKT and VFT scalings. The excellent fits of the experimental data to the VFT scalings confirm that the nanoperforated samples are bulk superinsulators. Also shown are the predicted $T^{\rm th}_{\rm dev}$ temperatures where finite-size effects are expected to cut-off the VFT fits for infinite samples and the corresponding experimental $T^{\rm exp}_{\rm dev}$ deviation temperatures . Note, again, the excellent match to the experimental data.  } \label{fig_GT_VFT}
	\end{center}
\end{figure}

In Figure\,\ref{fig_GT_VFT} we present the conductance dependencies on temperature for the two perforated films, obtained in two different dilution refrigerators. The initial film is a superconductor, with the critical transition temperature $T_{\rm C} = 1.8$K. The conductance of the perforated films, on the contrary, shows superinsulating behaviour by approaching 0 instead of diverging. At low temperatures, noise is observed due to the current measurement circuit approaching its limit. The conductance of superinsulators completely vanishes below a finite critical temperature, see Methods. The precise value of the critical temperature is determined by a BKT scaling in 2D- and by a VFT scaling in 3D systems\,\cite{vft}, with the two behaviours differing by their distinct powers in the exponential scaling law, see Methods for details. Although the width of the perforated films is only 10 nm, the 3D VFT conductance scalings with transition temperatures $T_{\rm VFT} = 0.31$K for the 2mm long sample and $T_{\rm VFT} = 0.275$K for the 1 mm sample are clearly favoured, indicating that these are bulk superinsulators.

As in the case of superconducting films, the scaling is of course cut off by the finite size effects\,\cite{benfatto, schneider}. Due to the higher power in the scaling exponent though, these finite size effects are more important for 3D systems than for 2D ones. The finite size effects can be estimated by cutting off the exponential increase of the correlation length when approaching the transition at the smallest dimension of the sample, in this case 0.1 mm, see Methods. The so obtained estimates predict deviations from the VFT scalings at a temperature $T_{\rm dev} \approx 0.61$K for the 2 mm film and at a temperature $T_{\rm dev} \approx 0.55$K for the 1 mm film, which are again in excellent agreement with what is measured in the experiment.

\begin{figure}
	\begin{center}
		\includegraphics[width=\linewidth]{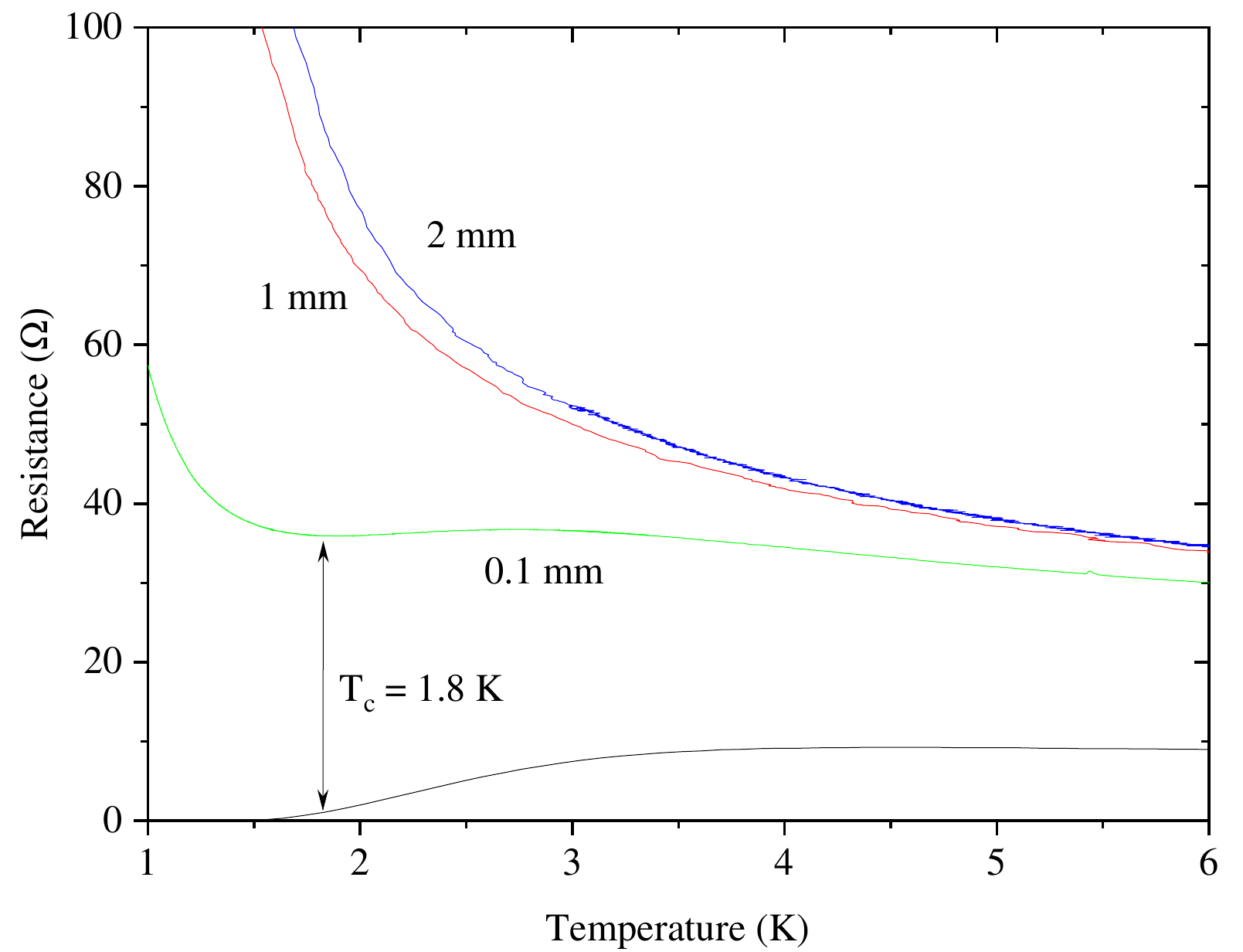}	
		\caption{\textbf{Temperature dependence of the resistance of a 0.1 mm small film.} Temperature dependence of the resistance of a 0.1 mm nanopatterned  film, showing a residual dip in correspondence of the superconducting critical temperature of the original, non-patterned film. }  \label{fig_add}
	\end{center}
\end{figure}

In large samples, showing the infinite-system behaviour, all vestiges of superconductivity are lost, since the system is in a different, superinsulating phase. When the sample size is decreased and finite-size effects become more important, though one can still observe a dip in the resistance at the original superconducting temperature $T= 1.8 $K. This is shown in Figure\,\ref{fig_add} for a 0.1 mm nanopatterned film.

\begin{figure}
	\begin{center}
		\includegraphics[width=\linewidth]{%Fig_RvsB_diff_T.pdf
			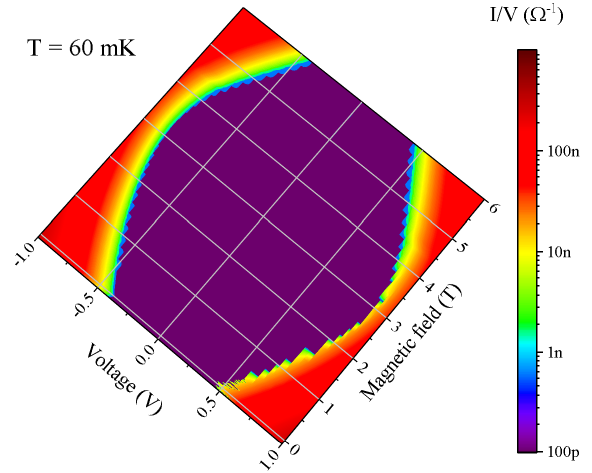
		}	
		\caption{\textbf{Conductance in the superinsulating state.} Magnetic field and voltage dependence of the conductance of the nanoperforated sample, measured at $T=60$ mK, below the superinsulating transition.} \label{fig_con}
	\end{center}
\end{figure}

\begin{figure}
	\begin{center}
		\includegraphics[width=\linewidth]{%Fig_RvsB_diff_T.pdf
			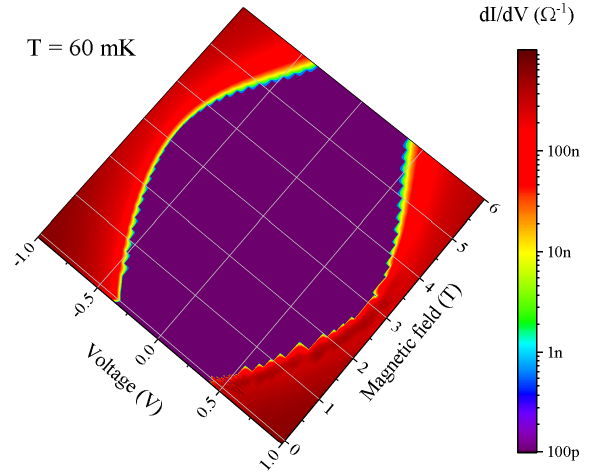
		}	
		\caption{\textbf{Differential conductance in the superinsulating state.} Magnetic field and voltage dependence of the differential conductance of the nanoperforated sample, measured at $T=60$ mK, below the superinsulating transition.} \label{fig_diffcon}
	\end{center}
\end{figure}

Below the VFT transition temperature, the nanoperforated system is in its superinsulating phase. In Figures\,\ref{fig_con} and \ref{fig_diffcon} we show the absolute and differential conductances measured at $T=60$ mK, deep in the superinsulating state. This confirms the vanishing of the conductance independently of the applied voltage (and magnetic field) in a range $|V| < V_{\rm cr} (B)$. This is the electric Meissner state of the superinsulator.

%
\iffalse

\begin{figure}
	\begin{center}
		\includegraphics[width=\linewidth]{Fig_RvsB_diff_T.pdf}	
		\caption{\textbf{Magnetoresistance.} Magnetic field dependence of resistance of nanoperforated film with length 2 mm at temperatures 300 and 700 mK}\,\label{fig_RvsB}
	\end{center}
\end{figure}

In Figure\,\ref{fig_RvsB} we present evolution of insulating properties of perforated film in magnetic field. With increasing magnetic field firstly resistance of perforated film increases, than reached maxima at $B = 2.5$T, and decreases after it. Increasing of temperature leads to a decrease in resistance, but the qualitative behavior of the magnetoresistance does not change.

\fi
%
Electric fields do not penetrate a superinsulator in its electric Meissner state. However, contrary to the dual superconductors, small open electric flux tubes with $\pm$ charges at their ends can be excited\,\cite{pion, book} if the energy provided by the external fields is sufficient. These small strings, whose gap and dimension are characterized by their string tension, are the purely electric equivalent of strong interaction pions.

\begin{figure}
	\begin{center}
		\includegraphics[width=\linewidth]{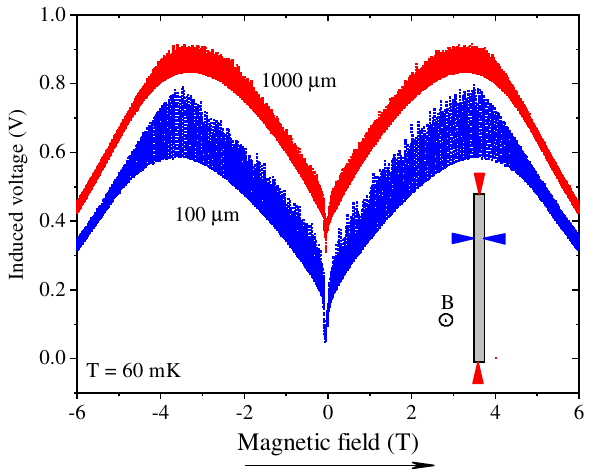}	
		\caption{\textbf{Polar nematic order induced by a magnetic field.} Magnetic field dependence of the induced voltage, measured at $T=60$ mK deep in the superinsulating state}\,\label{fig_VvsB}
	\end{center}
\end{figure}

When external electromagnetic fields $F^{\mu \nu}$ of a sufficient strength are applied to a superinsulating sample, they create a four-tensor current of strings $J^{\mu \nu}$, comprising, e.g., a vector density $J^{0i}$ for an electric field $E^i$ in the plane and a tensor current $J^{12}$ for a magnetic field $B^3$ perpendicular to it, see Methods. Let us consider first just a magnetic field $B^3=B$ in the third direction. When such a magnetic field is applied electric strings are excited. Due to their short-range antiferroelectric interaction, these strings tend to anti-align themselves along a nematic director ${\bf n}$. The external magnetic field, however, tends to push anti-aligned strings apart, see Methods. As a consequence, when the density of strings is not too high, this external interaction wins the competition and it is energetically favourable to have aligned strings moving all in the same direction under the influence of the external orthogonal magnetic field, thereby maintaining a large distance and a small interaction energy among them. The equilibrium configuration is thus a steady current of aligned electric strings, which amounts to a ferroelectric polar nematic order, detected as the polarization pattern shown in Fig.\,\ref{fig_VvsB}. When the magnetic field increases, however, the density of strings becomes large. In this case, the interstring interaction becomes the dominant one and an antiferroelectric nematic order is favoured, which does not give rise to polarization effects. The increase and subsequent decrease of the induced voltage by increasing the magnetic field is thus explained by the excitation of electric strings with a ferroelectric nematic order at low densities and an antiferroelectric nematic order at high densities.

Note that in zero magnetic field a small induced voltage is observed. We expect this to be caused by loops of electric fields in the ground state. The superinsulating ground state is a condensate of magnetic monopoles, see Methods, whose magnetic field becomes squeezed in the plane in the case of thin films. In this case, the monopoles become tunnelling events in which vortices appear and disappear in the ground state. Such events correspond to time-varying magnetic fields and are accompanied by electric field loops, essentially electric vortices. In the bulk of the sample these typically cancel out but they can give rise to polarization effects on the boundaries of finite systems when these are connected to measuring devices with a necessarily lower resistance than the sample. Note that this ground state effect is not affected by an applied constant magnetic field which does not create electric field loops. The only effect of a constant magnetic field is to excite electric strings. Of course, the excitation of a small amount of electric strings by the finite temperature can also not be excluded.

\begin{figure}
	\begin{center}
		\includegraphics[width=\linewidth]{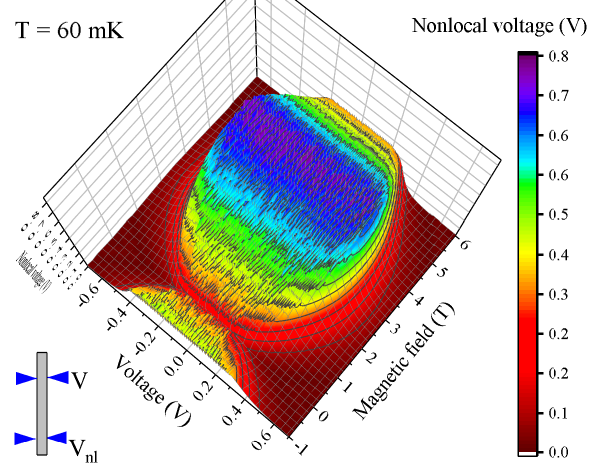}	
		\caption{\textbf{Induced non-local voltages.} Magnetic field and applied voltage dependencies of induced polarization voltages at different positions of applied voltage probes. Temperature of measurements is $T = 60$ mK. } \,\label{fig_nonloc_new1} 	\end{center}
\end{figure}

\begin{figure}
	\begin{center}
		\includegraphics[width=\linewidth]{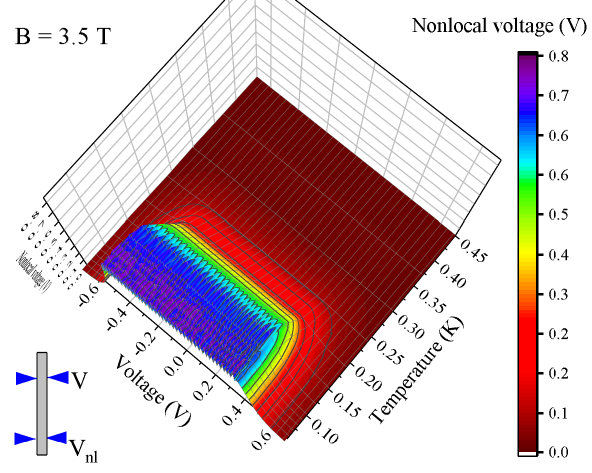}	
		\caption{\textbf{Induced non-local voltages.} Temperature and applied voltage dependencies of induced polarization voltages at different positions of applied voltage probes, measured with an applied perpendicular magnetic field of 3.5 T}\,\label{fig_nonloc_new2} 	\end{center}
\end{figure}

In Fig.\,\ref{fig_VvsB} we present the in-plane voltages induced by a magnetic field perpendicular to the perforated film at $T=60$ mK, deep in the superinsulating state.
When an additional voltage is applied in plane, one can measure the non-local induced voltage $V_{\rm nl}$ at a different location on the sample. These measurements are shown in
Fig.\,\ref{fig_nonloc_new1} and Fig.\,\ref{fig_nonloc_new2} as a function of magnetic field for a temperature of $T=60$ mK and of temperature for a magnetic field $B= 3.5$T. They
show that the polar nematic order vanishes when the applied voltage exceeds the critical value for superinsulation and confirm thus the origin of this phenomenon.

\begin{figure*}
	\begin{center}
		\includegraphics[width=\linewidth]{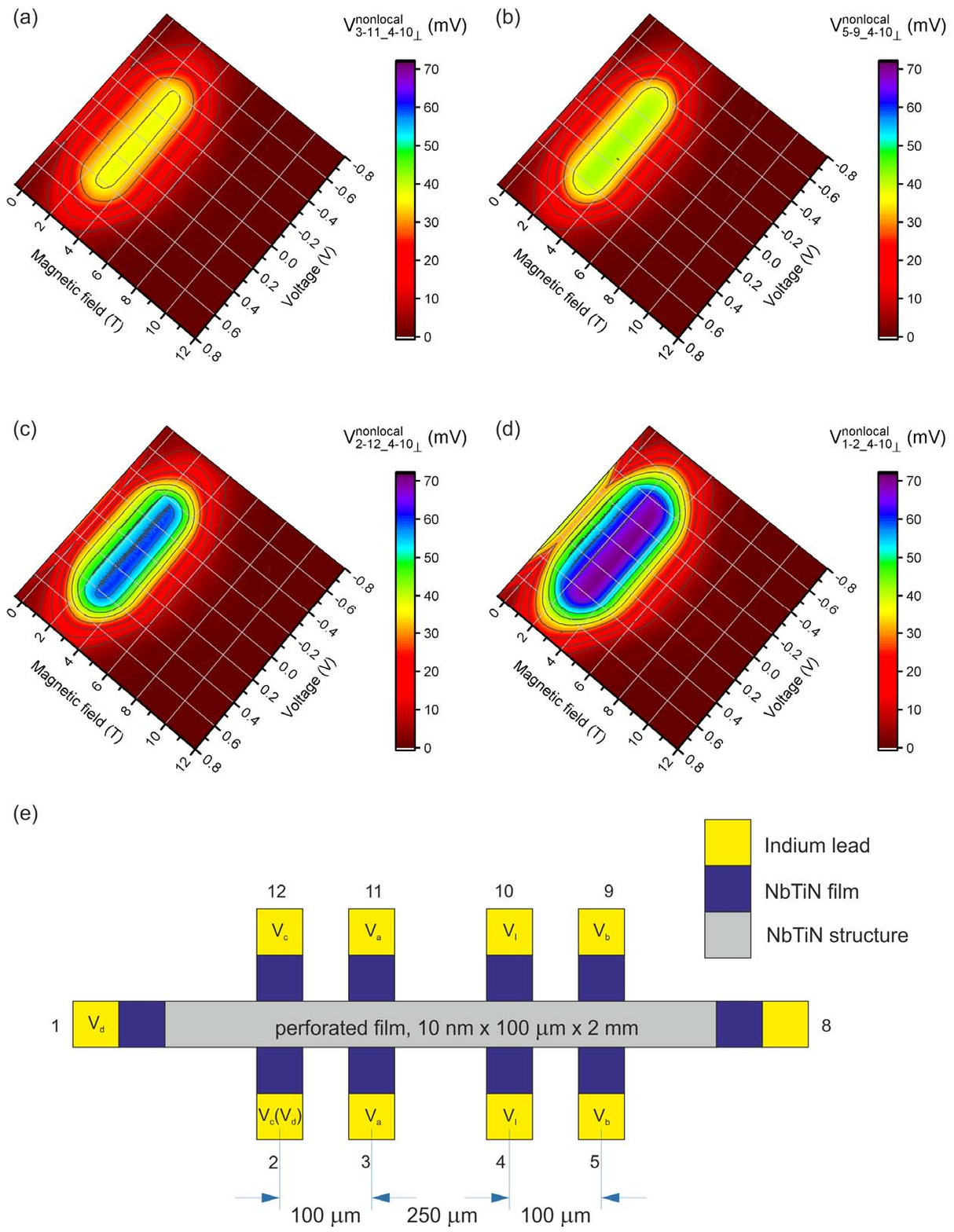}	
		\caption{\textbf{Induced non-local voltages.} (a-d) Magnetic field and applied voltage dependencies of induced polarization voltages at different positions of voltage probes, indicated by the first subscripts on the induced non-local voltages. Applied voltage contacts are indicated by the second subscripts on the induced non-local voltages. Last number of second subscript is ground. Temperature of measurements is $T = 160 mK$. (e) Schematic measuring circuit.}\,\label{fig_nonloc_V} 	\end{center}
\end{figure*}

\begin{figure*}
	\begin{center}
		\includegraphics[width=\linewidth]{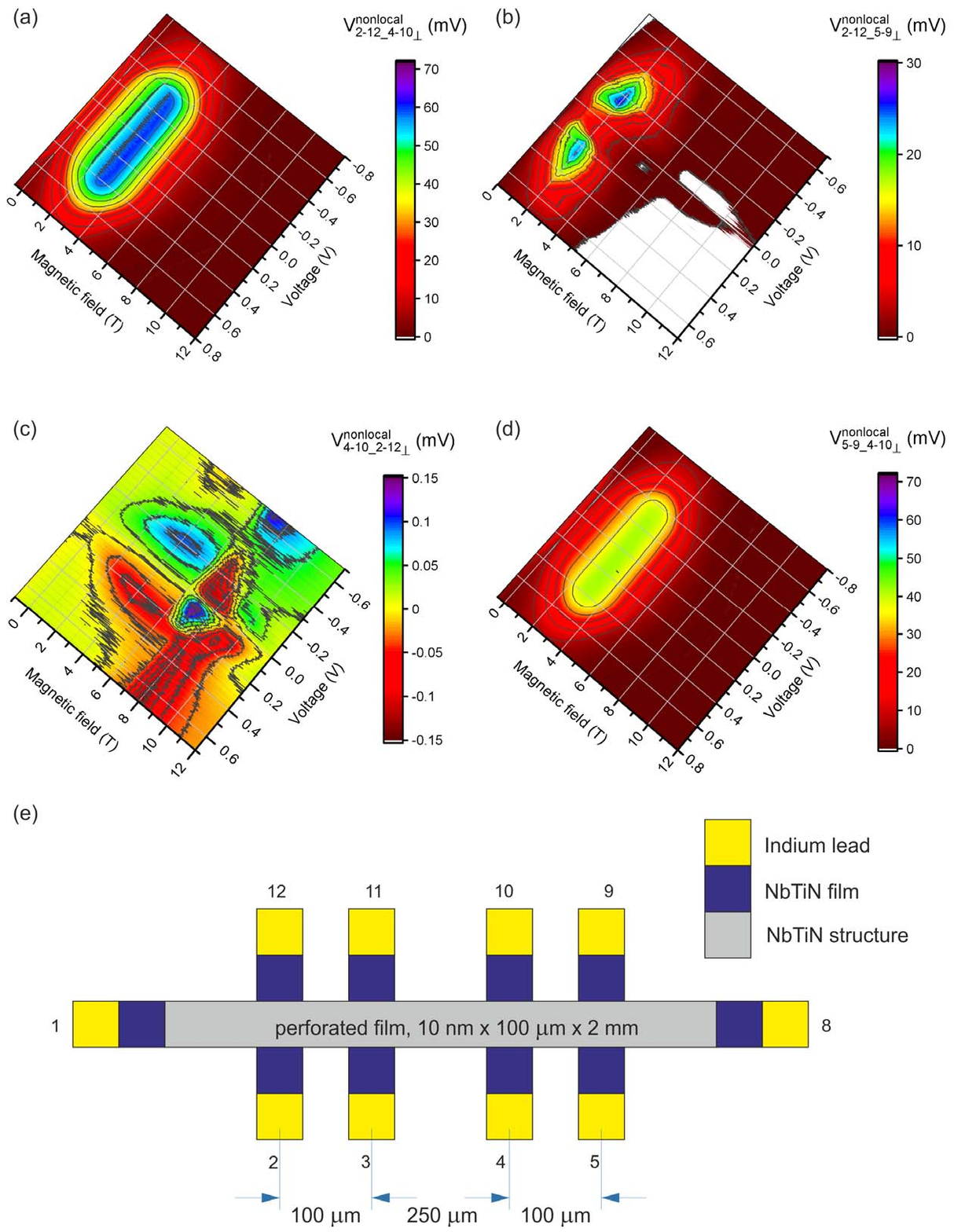}	
		\caption{\textbf{Induced non-local voltages.} (a-d) Magnetic field and applied voltage dependencies of induced polarization voltages at different positions of voltage probes, indicated by the first subscripts on the induced non-local voltages. Applied voltage contacts are indicated by the second subscripts on the induced non-local voltages. Last number of second subscript is ground. Temperature of measurements is $T = 160 mK$. (e) Schematic measuring circuit. }\,\label{fig_nonloc_VI} 	\end{center}
\end{figure*}

When both a perpendicular magnetic field and an in-plane electric field are present, complex polarization patterns can be detected in the film plane, depending on where the voltage is applied and where the non-local voltage is measured. In Figure\,\ref{fig_nonloc_V} and Figure\,\ref{fig_nonloc_VI} we present such non-local polarization patterns.  The first couple of subscripts on the applied voltage $V$ indicates the probes, the second couple the contacts where external voltage is applied. These measurements were done at a temperature $T=160 m$K, below the superinsulating transition.

Note that in all these measurements a magnetic field of at least approximately 2-3 T is required to produce appreciable induced voltages. Since the length scale $\ell$ associated with a magnetic field $B$, in a model where $2e$ is taken as the unit charge, is given by
\begin{equation}
\ell = \sqrt{\hbar \over 2eB} = {18 \over \sqrt{B(Tesla)}} nm \ ,
\label{length}
\end{equation}
this means that, in this sample, the electric pion excitations have a dimension smaller than $\approx 13-10 nm$, which is essentially the width of the perforated films, see Methods. The fundamental string excitations of the model, thus, feel the third dimension and this is why these perforated films show 3D bulk superinsulation.

\iffalse

\begin{figure*}
	\begin{center}
		\includegraphics[width=\linewidth]{fig_Wire.pdf}	
		\caption{\textbf{Electric field induced by current.} (a) Real part of voltage induced to wire around sample by current flow through superinsulator in different magnetic filed. (b) Imaginary part of voltage induced to wire around sample by current flow through superinsulator in different magnetic filed. (c) AC current flow through superinsulator in different magnetic filed. AC voltage is 100 mV.  Temperature of measurements $T = 160 mK$. (d) Schematic measuring circuit. The wavy line shows the wire around the sample.}\,\label{fig_Wire} 	\end{center}
\end{figure*}

\fi

\section*{Methods}~~
\subsection{Magnetic monopoles and superinsulation}
Superinsulation occurs in granular materials, where the granularity can be either emergent\,\cite{granular1, granular2} or synthetic as in JJA\,\cite{jja} or in the present nanopatterned metamaterial. Whatever case is realized, such a system can be modelled as an effective JJA with random couplings. One might thus be lead to think that disorder-induced localization is the origin of the vanishing conductance. This has been already disproved by several experiments, even on films with irregular granularity \cite{pion, mironov2}. In the present nanopatterned material, however, the granularity is perfectly regular and ordered. If disorder would be the root cause of the vanishing conductance, this phenomenon should not take place in these nanopatterned materials. But it clearly does! This unambiguously excludes disorder as the origin of superinsulation.

The point is that the circulation of the local order parameter phase on neighbouring granules can lead to the formation of coreless vortices and that there is nothing preventing these vortices to end inside the system so that their ends form a magnetic monopole-antimonopole pair. Such configurations must be included in the electromagnetic effective action because the gauge group in a lattice-like granular system is compact\,\cite{polyakov}. The presence or absence of magnetic monopoles becomes, thus, only a dynamical issue depending on the value of the effective coupling constant inside the material\,\cite{moncond}. And these monopoles can condense, leading to superinsulation as electric confinement, without any role of disorder.

The compact Euclidean effective action determining the properties of the granular material follows from the standard general relation and is (we use natural units $c=1$, $\hbar = 1$, $\varepsilon_0=1$)
\begin{eqnarray}
	&&Z = \sum_{ \{ m_{\mu} \} } \int {\cal D} A_{\rm\mu}\  {\rm e}^{-S} \ ,
	\nonumber \\
	&&S = \sum_x {1\over 4f^2} \left( F_{\rm{\mu \nu}} -2\pi S_{\mu \nu} \right)^2  \ ,
	\label{compactqed}
\end{eqnarray}
where $x$ denotes the granule sites, $F_{\mu \nu}= d_{\mu} A_{\nu} - d_{\nu} A_{\mu}$ is the usual electromagnetic field tensor, the integers $m_{\mu} = (1/2) K_{\mu \alpha \beta} S_{\alpha \beta}$ describe the conserved magnetic monopole currents, $K_{\mu \alpha \beta}$ being the proper gauge-invariant lattice version of the curl operator $\epsilon_{\mu \nu \alpha \beta} \partial_{\nu}$ (see e.g.\,\cite{book}) and the dimensionless coupling $f$ is the effective strength of the Coulomb interaction in the material. The integer magnetic currents $m_{\mu}$ are the only physical degrees of freedom contained in the plaquette integers $S_{\mu \nu}$. Non-relativistic effects can be incorporated by an effective speed of light smaller than $c$. If the unit of an electric charge is $2e$, magnetic charge is quantized in units of $2\pi/2e = \pi/e$. For simplicity, we have presented here the partition function at zero temperature; a finite temperature can be incorporated by considering a finite lattice of length $1/k_BT$ in the Euclidean time direction, with periodic boundary conditions for our bosonic fields, see e.g.\,\cite{book}.

Integrating out the gauge field $A_{\mu}$ one obtains the model formulated solely in terms of magnetic monopole currents
\begin{eqnarray}
&&Z= \ Z_0 \cdot \sum_{ \{ m_{\mu} \} } {\rm e}^{-S} \ ,
\nonumber \\
&&S= \  {\pi^2 \over f^2} \sum_x m_{\mu} {1\over -\nabla^2} m_{\mu} \ ,
\label{charmon}
\end{eqnarray}
where $Z_0$ is the partition function of the non-compact model, describing normal Mott insulators, and $-1/\nabla^2$ is a short-hand notation for the Poisson Green's function $G({\bf x})$, defined by
\begin{equation}
-\nabla^2 G({\bf x}) = \delta^2 ({\bf x} ) \ .
\label{poisgre}
\end{equation}
Therefore monopoles interact via a Coulomb potential in a 4D Euclidean space. For weak coupling they are suppressed and we have a Mott insulator, for strong coupling they Bose condense\,\cite{polyakov}, see\,\cite{book} for a review. In this state, electric charges are confined by a linear potential generated by an electric flux tube between the charges and holes\,\cite{confinement, book} and we have a superinsulator.
\vspace{0.2cm}

\subsection{Vogel-Fulcher-Tamman scaling}
Superinsulators turn into Mott insulators when the temperature is raised enough, so that the the formation of a condensate of monopoles is suppressed. In 2D, this is a BKT transition\,\cite{ber, kt1, kt2}, characterized by the BKT scaling of the resistance
\begin{equation}
R(T) \propto {\rm e}^{b_2\over \sqrt{|T-T_{\rm BKT}|}} \ .
\label{bkttr}
\end{equation}
Below this transition, the resistance is infinite, and the system, therefore, is a dual mirror of superconductors, which demonstrate a zero resistance below this BKT transition. This is what distinguishes superinsulators from Mott insulators: in superinsulators the resistance is infinite in an entire finite temperature domain. In 3D, the physics is the same but the resistance scaling is different \cite{vft},
\begin{equation}
R(T) \propto {\rm e}^{b_3\over |T-T_{\rm VFT}|} \ .
\label{vfttr}
\end{equation}
This behavior is known as a Vogel-Fulcher-Tamman (VFT) scaling and is the hallmark of 3D bulk superinsulators.

If the voltage applied to a superinsulator is below a critical threshold voltage $V_{\rm c1}$, the current does not flow across the system. A current can be passed through a superinsulator by applying a voltage above a critical threshold voltage $V_{\rm c1}$. In this case, the electric flux tubes connecting the charges of the opposite signs and preventing their further separation are stretched to the extent that they touch the boundary of the sample. Monopoles are still at work but the flux tube is long enough so that current can pass through the system. This is the electric mixed state\,\cite{pion}, the state dual to the magnetic mixed state in type-II superconductors, in which the applied magnetic field is above a critical value $H_{\rm c1}$ and penetrates the sample in a form of Abrikosov vortices. Analogously, in superinsulators, the electric fields are expelled entirely from superinsulators as long as applied voltages remain below $V_{\rm c1}$, we refer to this phenomenon as to the\,\textit{electric} formation of a Meissner state\,\cite{pion, book}. At the intermediate voltages, $V_{\rm c1}<V<V_{\rm c2}$, electric strings form within a superinsulator. And, finally, when the applied voltage is above a second critical voltage $V_{\rm c2}$, superinsulation is destroyed entirely.
\vspace{0.2cm}

\subsection{Electric strings and their equations of motion}
In the superinsulating state, the fundamental excitations are strings, representing thin electric flux tubes  see\,\cite{book} for a review. Contrary to Abrikosov vortices in superconductors, these strings are open, with a Cooper pair and a Cooper hole at their ends. Their gap and dimension are governed by the string tension $\sigma$, which is the string equivalent of mass for particles. Such open electric flux tubes can be thought of as the electric equivalent of strong interaction pions.

As the dynamics of charged particles is encoded in a four-vector current $j^{\mu}$, describing their world-lines and coupled to a vector gauge potential $A_{\mu}$, the dynamics of strings is encoded in an antisymmetric tensor current $J^{\mu \nu}$, describing their world-sheets and coupled directly to the electromagnetic fields $F_{\mu \nu}$, which are the fundamental fields\,\cite{pion, book}; there is no more a vector potential, except to describe the coupling of the string endpoints, described again by a current $j^{\mu}=\partial_{\nu}J^{\mu \nu}$ encoding the world-sheet boundaries.

The equation of motion for strings in the superinsulating state is given by
\begin{equation}
J^{\mu \nu} \propto F^{\mu\nu}  + \dots \ ,
\label{ellon}
\end{equation}
which is the electric equivalent of the London equations $j^{\mu} \propto A^{\mu}$ for a superconductor (as in this case the $\dots$ denote higher-order in derivatives terms) \cite{pion, book}. The charged endpoints follow the usual electromagnetic laws.

When an applied external magnetic field $B^3 = B$ becomes sufficiently strong to excite one pion per unit volume, $|B| > O(\sigma)$, this creates a tensor current $J^{12}$, which can be understood as the current in direction 1 of a unit vector charge in direction 2 minus the current in direction 2 of a unit vector current in direction 1, where the vector charge represents the electric field. A rigid string thus moves as a whole in the direction ${\bf E} \times {\bf B}$ and, correspondingly, the Lorentz force on the endpoint charges stretches the string in the direction of ${\bf E}$, making it even more rigid. Correspondingly, when a sufficiently strong electric field $E^i$, $i=1,2$ is applied in the plane perpendicular to the magnetic field, this creates static strings with vector density $J^{0i}$ in the interior of the sample\,\cite{pion}.

Electric strings interact with each other via a short-range, Yukawa-type potential\,\cite{book}. As for magnetic flux tubes in supercoundcutors, parallel electric strings of the same polarity repel each other, while electric strings of opposite polarity attract each other.

\vspace{0.2cm}

\subsection{Finite-size effects}
As usual, phase transitions in real samples are broadened due to finite-size effects. We can estimate the temperature where these deviations start to show up by considering the correlation length of the VFT transition. This is given by\,\cite{vft, book}
\begin{equation}
\xi  = \xi( T=\infty) \ {\rm e}^{b_3\over |T-T_{\rm VFT}|} \ ,
\label{corrvft}
\end{equation}
where $\xi(T=\infty)\approx\ell$, with $\ell$ being the ultraviolet cutoff, which in our case is the effective lattice spacing of the nanopatterned material\,\cite{book}. At the transition, the correlation length diverges exponentially, in analogy to the 2D BKT case but with a different power in the exponent. However, this divergence is clearly cut off at the sample size $L$ in real samples. One can estimate the temperature at which this deviation shows up by setting $\xi = L$ in the above equation. This gives
\begin{equation}
T_{\rm dev} = T_{\rm VFT} + {b_3 \over {\rm ln} (L/\ell) } \ .
\label{dev}
\end{equation}
For the 2 mm sample, $T_{\rm VFT}=0.31 ^{\circ}K$, $b_3=2.05 ^{\circ}K$, $L=0.1\ mm$, and $\ell\approx100\,nm$, which gives $T_{\rm dev}\approx0.61^{\circ}K$. For the 1 mm sample, $T_{\rm VFT}=0.275 ^{\circ}K$, $b_3=1.9 ^{\circ}K$, with $L=0.1\ mm$, and $\ell\approx100\,nm$, as for the other sample: this gives $T_{\rm dev}\approx0.55^{\circ}K$.
Note that, due to the larger power in the exponential scaling, finite-size effects are much more important for VFT than for BKT scaling.

	\section* {Acknowledgements}
	The work by V.M.V. was supported by Terra Quantum.
	\section* {Author contribution} A.Yu.M., M.C.D., C.A.T., and V.M.V. conceived the work, A.Yu.M. and D.A.N. did the experiments,  M.C.D., C.A.T., and V.M.V. carried out calculations, all authors discussed results and wrote the paper.
	\section*{Competing interests}\,The authors declare that they have no competing interests.
	\section* {Correspondence and requests for materials} should be addressed to V.M.V. (vv@terraquantum.swiss).

%\newpage ~~~\\

%\newpage ~~~\\

\end{document}